\documentclass[11pt]{elsarticle}

\usepackage[margin=0.8in]{geometry} 
\usepackage{multicol}

\usepackage{epsfig}
\usepackage{latexsym}
\usepackage{epic}
\usepackage[leqno]{amsmath}
\usepackage{amssymb}
\usepackage{url}
\usepackage[all]{xy}

\newtheorem{lemma}{Lemma}[section]
\newtheorem{theorem}{Theorem}
\newtheorem{proposition}[lemma]{Proposition}

\newcommand{\Rr}{\mbox{$\mathcal R$}}

 \newcommand{\Rc}{\mathcal R}

\newcommand{\Cc}{\mathcal C}

\newcommand{\Ww}{\mathcal W}

\newcommand{\Bb}{\mathcal B}
\newcommand{\Aa}{\mathcal A}

\newcommand{\Hh}{\mbox{$\mathcal H$}}

\newcommand{\R}{{\mathbb R}}
\newcommand{\N}{{\mathbb N}}
\newcommand{\pf}{\noindent{\em Proof: }}
\newcommand{\epf}{\hfill\hbox{\rule{3pt}{6pt}}\\}

\newcommand{\Max}{{\rm Max}}

\newcommand{\Min}{{\rm Min}}
\newcommand{\Mm}{{\mathcal M}}

\newcommand{\ra}{\rightarrow}
\newcommand{\Ra}{\Rightarrow}

\newcommand{\CC}{{\mathcal C}}

\newcommand{\ext}{{\rm ext}}

\newcommand{\0}{\emptyset}

\newcommand{\wpg}[1]{[{#1}]_{wk} }
\newcommand{\spg}[1]{[{#1}]_{st} }
\newcommand{\pg}[1]{[{#1}]_{pt} }
\newcommand{\y}[1]{[{#1}]_{yy}}

\begin{document}

\title
{On Patchworks and Hierarchies}

\author{Andreas Dress$^1$}
\address{$^1$CAS-MPG Partner Institute for Computational Biology, Shanghai Institutes for Biological Sciences, Chinese Academy of Sciences, 320 Yue Yang Road, Shanghai, China, and MiS, Leipzig}
\author{Vincent Moulton$^2$}
\address{$^2$School of Computing Sciences, University of East Anglia, Norwich, NR4 7TJ, UK}
\author{Mike Steel$^3$}
\address{$^3$Biomathematics Research Centre, Department of Mathematics and Statistics, University of Canterbury, Private Bag 4800, Christchurch, New Zealand}
\author{Taoyang Wu$^4$}
\address{$^4$Department of Mathematics, National University of Singapore, Block S17, 10, Lower Kent Ridge Road, Singapore 119076}


\date{}

\begin{abstract}
Motivated by questions in biological classification, we
discuss some elementary combinatorial and computational properties of
certain set systems that generalize hierarchies,
namely, `patchworks', `weak patchworks', `ample patchworks' and  `saturated patchworks'
and also outline
how these concepts relate to an
apparently new `duality theory' for cluster systems
that is based on the fundamental concept of `compatibility' of clusters.
\end{abstract}
\maketitle


\section{Introduction}

In various fields of classification, such as evolutionary biology, a nested hierarchy is regarded as the ideal object for describing the relationships between the objects under consideration
(e.g. species).  Such a hierarchy corresponds to the `clades' of a rooted tree, and the notion in taxonomy traces back at least to Linnaeus \cite{lin}, if not Aristotle \cite{arist}, that is, it was considered years before the concept of an evolutionary phylogenetic tree was formed.

But set systems that are more complex than hierarchies are also frequently  relevant  in this setting for two reasons:
Firstly, even when the underlying structure can be assumed to be `tree-like', the data itself may not be, and so the question of whether there are canonical ways to construct a hierarchy from an arbitrary set system (or other structure, such as a graph representing, for example, a genealogy,  or a topological space)
arises.  This topic has been explored recently  \cite{dre}, \cite{dre2}, and we briefly consider some further aspects of it in this note.

Our main interest, however, lies in the second reason for  dealing with more complex set systems than hierarchies; namely to accommodate settings where a tree does not provide an entirely accurate classification. In the case of evolutionary biology, for instance, processes such as lateral gene transfer, and the formation of hybrid species, give rise to non-treelike evolutionary histories  (see, for example, \cite{doo}, \cite{lin}, \cite{hus}).

Two distinct approaches have been developed for
accommodating non-tree like processes.
One  approach attempts to use overtly graph-theoretic approaches to
construct different types of directed acyclic graphs that
could be appropriate for describing a `network
of life'.  The other is to simply
consider set systems on taxa that can arise from such networks  (e.g. via their so-called `soft-wired' or `hard-wired' clusters \cite{hus}) or -- more directly -- from the complex pattern of the presence and absence of genetic markers across taxa.  For this second approach,  some relaxations of hierarchies have been developed, such as the notion of weak hierarchies, pyramids, or $k$-compatible set systems (see, for example,   \cite{dre4}, \cite{mir}, \cite{sem}).

Here, we explore a different type of relaxation, {\em viz.}~various types of `patchworks' -- a class of set systems that was introduced into phylogenetics in \cite{BDM99}.  These set systems can be generated by iteratively
enlarging a given `generating set' (Lemma \ref{lem:ext:gss} below) and, when the resulting  set system is sufficiently abundant, it will harbour at least one fully resolved or `maximal' hierarchy  ({\em cf.}~Theorem~\ref{motive}).
Patchworks also turn up naturally in the context of a certain `duality theory' for cluster systems and can be used to associate a canonical hierarchy with an arbitrary set system. Patchworks may thus provide a new tool for studying evolutionary relationships in a setting where hierarchies can be obscured by reticulate processes such as extensive lateral gene transfer, and
for analysing collections of subsets of taxa created according to the presence/absence patterns
of genetic loci thereby providing some insight into  the extent to which a single
tree (as opposed to a more complex network) may describe
the evolution of the taxa  \cite{dag}. \\

The paper is organized as follows: We introduce some basic definitions and notation in the next section, and the main concepts we are going to investigate in Section \ref{patchworks}. We collect some relevant examples in Section \ref{examples}, and then present one of our two main results, {\em viz.} Theorem~\ref{motive}, in the next section. Then, we discuss a duality theory for cluster systems in Section \ref{dual} (containing the other main result --  Theorem~\ref{main} --  characterizing `self-adjoint' cluster systems)
as well as certain closure operators for cluster systems in Section \ref{generators}.
In Section \ref{real-world}, we illustrate these concepts
using a biological data set, the so-called `Belgian Transmission Chain' (of the human immunodeficiency virus (HIV), {\em cf.} \cite{lem}).
We conclude the paper with a collection of comments and remarks regarding possible applications, extensions, and some open questions (Section \ref{conclusion}).

\section{Basic definitions and notation}\label{basic}
We begin with some terminology and definitions. Let $X$ be a finite set of cardinality $n$
at least $2$ that we fix once and for all.

The partially ordered set consisting of all non-empty subsets of $X$  (ordered by set inclusion `$\subseteq$') whether empty or not is denoted by $\Cc(X)$, every subset $\Cc$ of $\Cc(X)$ is called
a {\em cluster system} (for $X$), and the set of all cluster systems $\Cc\subseteq  \Cc(X)$ is denoted by $\Cc^{(2)}(X)$.   Further,  given any cluster system $\Cc\in  \Cc^{(2)}(X)$, 
we denote 
\begin{itemize}
\item[--] 
%
by $\bigcup \Cc:=\bigcup_{C\in \Cc}C$ the union and by  
$\bigcap \Cc:=\bigcap_{C\in \Cc}C$ the intersection of all clusters $C\in \Cc$,
\item[--] 
by 
$\Max(\Cc)$ the set 
$\{C\in \Cc:\{C'\in \Cc: C \subsetneq C' \}=\0\}$ of all maximal clusters in $\Cc$,
\item[--] 
by 
$\Min(\Cc)$ the set 
$\{C\in \Cc:\{C'\in \Cc: C' \subsetneq C\}=\0\}$ of 
all minimal clusters in $\Cc$, 
\end{itemize}
and
given -- in addition -- any cluster $A\in \Cc(X)$, we denote
\begin{itemize}
\item[--]  
 by $\Cc(\subseteq\!A):=\{C\in \Cc: C \subseteq A\}$ the 
 %
 cluster system  consisting of all sub-clusters of $A$ in $\Cc$, 
 
 \item[--] by $\Cc(\supseteq\!A):=\{C\in \Cc: C \supseteq A\}$ the
 %
 cluster system consisting of all clusters in $\Cc$ containing $A$, 
\item[--]  
and we put $\Max(\Cc\!:\!A):= \Max\big(\Cc(\subseteq\!A)-\{A\}\big)$.
\end{itemize}

Two subsets $A$ and $B$ of $X$ are called {\em compatible} -- which we denote by `$A \!\!\parallel \!\!B$' -- if and only if $A\cap B \in \{\emptyset, A, B\}$ holds; otherwise $A$ and $B$ are said to be {\em incompatible} -- which we denote by `$A \!\nparallel\!B$'. Two cluster systems  $\Aa,\Bb\in  \Cc^{(2)}(X)$ are called {\em compatible} if $A\!\!\parallel \!\!B$ holds for every $A\in \Aa$ and $B\in \Bb$, which we denote by `$\Aa \!\!\parallel \!\!\Bb$'.
And, given any cluster system $\Cc\in  \Cc^{(2)}(X)$,
\begin{itemize}
\item[--] we denote by $\Cc_{extr}$ the collection of
those clusters $A\in \Cc$ for which no clusters $B,C\in \Cc$  with $A = B \cup C$ and $B \!\nparallel\!C$ exist,
(i.e., the collection of clusters in $\Cc$ that are
{\em extremal} relative to the
`weak patchwork closure operation' to be introduced below),
\item[--] we denote by $\Cc^*$ the largest cluster system in $ \Cc^{(2)}(X)$ that is compatible with $\Cc$, i.e., the collection of all non-empty subsets $A$ of  $X$ that are compatible with every
cluster $C\in\Cc$ --
this cluster system will also be called the {\em adjoint} of $\Cc$,
\item[--] and we denote
the {\em double adjoint} $(\Cc^*)^*$ of $\Cc$, i.e., the adjoint of the adjoint of $\Cc$, simply by $\Cc^{**}$,
\item[--] while the intersection  $\Cc^*\cap \Cc^{**}$ will be denoted by $\Cc_\circ$.
\end{itemize}

Finally, the {\em Hasse diagram} $H_{\Cc}$ associated with a  cluster system $\Cc \in  \Cc^{(2)}(X)$ is the directed graph with vertex set $\Cc$ and arc set $$E_{\Cc}:=\{(C_2,C_1)\in \Cc^2: |\{C\in \Cc: C_2 \subseteq C \subseteq C_1\}|=2\}.$$

\section{Patchworks}\label{patchworks}

\medskip

In many applications, it is useful to consider
families of cluster
systems that satisfy additional constraints.
{\em Hierarchies}, studied in phylogenetics as well as in combinatorial optimization (where they are dubbed {\em laminar families} \cite{kor}),
form a particularly  popular example.  Formally, a cluster system $\Cc\in  \Cc^{(2)}(X)$ is a called a (generalized) {\em hierarchy on
$X$} -- or, for short, an {\em $X$-hierarchy} or even just a {\em hierarchy} -- if any two clusters $A,B\in \Cc$ are compatible or, equivalently, if  $\Cc\subseteq \Cc^*$ holds.
Further, a cluster system
$\Cc\in \Cc^{(2)}(X)$ is called {\em ample} if $C_1-C_2\in \Cc$ holds for every arc $(C_2,C_1)\in E_{\Cc}$.

Motivated by their applications in phylogenetics \cite{dre4,sem}, patchworks were
introduced in~\cite{BD-01,BDM99}
as cluster systems $\Cc\in  \Cc^{(2)}(X)$ for which the following holds:
\begin{equation}
\label{def:eq:wp}
A,B\in \Cc \text{ and }A \!\nparallel\!B\Ra A\cap B, A\cup B \in \Cc.
\end{equation}
Here, we will also be interested in related notions: A cluster system $\Cc\in \Cc^{(2)}(X)$ shall be called a {\em weak patchwork} if it only
satisfies the second half of the patchwork condition (\ref{def:eq:wp}):
\begin{equation}
\label{def:eq:wp2}
A,B\in \Cc \text{ and }A \!\nparallel\!B\Ra A\cup B \in \Cc.
\end{equation}
And it will be called a {\em saturated patchwork} if, given any two clusters $A,B\in \Cc$ with $A\!\nparallel\!B$,
the five clusters  $A\cap B, A\cup B, A-B,B-A$ and $ A\cup B - A\cap B$ are also contained in $\Cc$.
In particular, any saturated patchwork is a patchwork, and hence is also a weak patchwork. It is also obvious that a cluster system $\Cc\in  \Cc^{(2)}(X) $ is a saturated patchwork if and only if, given any three disjoint clusters $C_0,C_1,C_2\in \Cc(X)$ with $C_0 \cup C_1, C_0 \cup C_2\in \Cc$,
one also has $C_0,C_1,C_2,  C_1 \cup C_2, C_0 \cup C_1\cup C_2\in \Cc$.

\bigskip
The following simple facts regarding such cluster systems are easy to check and/or well-known:
\begin{enumerate}
\item[(F0)] Every hierarchy is a saturated patchwork and, so, in particular, is a patchwork, and every patchwork is a weak patchwork, while the converse
 does not hold unless one has $n= 2$.
\item[(F1)]
The map
$\Cc^{(2)}(X)\ra \Cc^{(2)}(X): \Cc \mapsto \Cc^*$
defines a {\em Galois connection}\footnote{see \url{http://en.wikipedia.org/wiki/Galois_connection}}
on $\Cc^{(2)}(X)$, i.e., one has $\Aa\subseteq \Aa^{**}$  as well as $$``\Aa \subseteq \Bb \Ra \Bb^*\subseteq \Aa^*  \Ra  \Aa^{**}\subseteq  \Bb^{**}\mbox{''}$$ for all cluster systems
$\Aa,\Bb\in  \Cc^{(2)}(X)$ and, thus, also
$$``\Cc=\Cc^{**}\iff\exists_{\Aa\in  {\Cc^{(2)}}(X)}  \Cc=\Aa^*\mbox{''}$$
as well as $(\Cc^{**})^*=(\Cc^*)^{**}=\Cc^*$
 for all
$\Cc\in  \Cc^{(2)}(X)$.

\item[(F2)]  One also has $$ ``\Aa \!\!\parallel \!\! \Bb \iff \Aa \subseteq \Bb^*
\iff \Bb^{**} \subseteq \Aa^*
\iff   \Bb^{**}\!\!\parallel \!\!\Aa   \iff \Aa^{**} \!\!\parallel \!\!  \Bb^{**}\mbox{''}$$
 as well as
$(\Aa \cup \Bb)^*= \Aa^* \cap \Bb^*$ and  $(\Aa \cap \Bb)^*\subseteq  \Aa^* \cup \Bb^*$ for
all cluster systems $\Aa,\Bb\in \Cc^{(2)}(X)$ (but not necessarily ``$\Aa \!\!\parallel \!\! \Bb \Ra \Aa^* \!\!\parallel \!\!  \Bb^*$\,'' as the example $X:=\{1,2,3\}$ and $\Aa:=\Bb:=\big\{\{1\}\big\}$ shows).

\item[(F3)]
The adjoint  $\Cc(X)^*$ of  $\Cc(X)$ is the cluster system $\Cc_{triv}(X):=\{C\subseteq X: |C|=1 \text{ or } C=X\}$ consisting of all {\em trivial} clusters in
$\Cc(X)$. It is
a hierarchy and is contained in $\Cc^*$ for every cluster system $\Cc \in  \Cc^{(2)}(X)$, and its union with any other hierarchy, weak patchwork, patchwork, or saturated patchwork $\Cc\subseteq \Cc(X)$ is also a hierarchy, a weak patchwork, a patchwork, or a saturated patchwork, respectively. In particular, one has $\big(\Cc \cup \Cc_{triv}(X)\big)^*=\Cc^*$ and, hence, also $\big(\Cc \cup \Cc_{triv}(X)\big)^{**}=\Cc^{**}$ for every cluster system $\Cc\in  \Cc^{(2)}(X)$.

More generally, given any cluster system
$\Cc\in  \Cc^{(2)}(X)$ for which there exists a partition $\Pi$  of a subset $A$ of $X$ such that $\Cc$ coincides with $\langle \Pi \rangle:= \{\bigcup \Pi' : \0\neq \Pi' \subsetneq \Pi\}$  (i.e.,  the disjoint union $\Cc\dot\cup\{\0,A\}$ is the `ring of sets' generated by $\Pi$, {\em cf.}~\cite{birk}), one has $\Cc^*\cap \CC=\Pi $ -- while $\Cc^*-\CC$ consists of all proper non-empty subsets of the sets in the partition $\Pi$ and the subsets of $X$ that either contain or are disjoint from $A$.

\item[(F4)]
The cardinality $|\Hh|$ of any $X$-hierarchy $\Hh$ never exceeds $2n-1$, and it coincides with this number if and only if $\Hh$ is a {\em maximal}  $X$-hierarchy if and only if $\Hh=\Hh^*$ holds.

\item[(F5)]
In particular, the following three assertions hold:
\begin{itemize}
  \item[(F5-i)] The adjoint $\Cc^*$ of any cluster system
$\Cc\subseteq  \Cc(X)$ that contains a maximal $X$-hierarchy $\Hh$ must be a  hierarchy
as it is necessarily contained in the adjoint $\Hh^*$ of  $\Hh$ which, however,  coincides with $\Hh$ itself.

  \item[(F5-ii)]
Conversely,
the adjoint $\Hh^*$ of any hierarchy $\Hh \subseteq  \Cc(X)$ contains a maximal $X$-hierarchy
as it is actually the union of all $X$-hierarchies and, hence, also of all maximal $X$-hierarchies that contain $\Hh$.
\end{itemize}
\begin{itemize}
  \item [(F5-iii)]  So, a cluster system $\Cc\subseteq  \Cc(X)$ is a hierarchy if and only if its adjoint contains a maximal $X$-hierarchy, in which case its double adjoint is also a hierarchy.
\end{itemize}

\item[(F6)] Also, an $X$-hierarchy $\Cc$ is a maximal $X$-hierarchy if and only if $X\in \Cc$ holds and either $(i)$ every cluster $C\in \Cc$ whose cardinality $|C|$ exceeds $1$ is the union of two proper (and, hence, necessarily disjoint) sub-clusters $C_1,C_2\in \Cc$ or, equivalently, $(ii)$ $\Cc$ is  ample and $ \{x\} \in \Cc$ holds for all $x \in X$ ({\em cf.} \cite{BD00} for generalizations regarding maximal $X$-hierarchies for infinite sets $X$).

\end{enumerate}

\begin{figure}[htb]
\centering
\includegraphics[width=13cm]{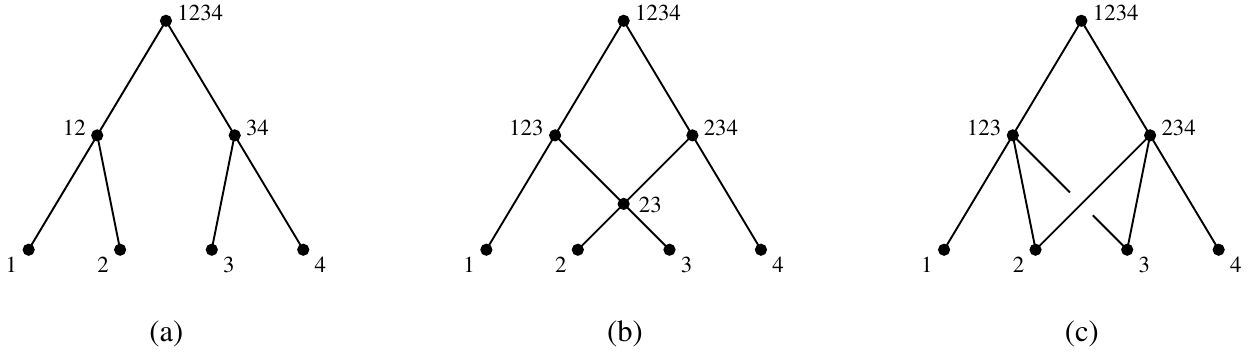}
\caption{The Hasse diagrams of (a) the hierarchy $\{1,2,3,4, 12,34, 1234\}$; (b) the patchwork $\{1,2,3,4, 23, 123, 234, 1234\}$; and (c) the weak patchwork $\{1,2,3,4,123, 234, 1234\}$.}
\label{fig1}
\end{figure}

Recall also (from, for example, \cite{edm}) that a cluster system $\Cc\in  \Cc^{(2)}(X)$ with
$X=\bigcup \Cc$ is a hierarchy if and only if
$\Cc(x) :=\bigcap \Cc(\supseteq \{x\}) \in \Cc$ holds for every $x\in X$ and the Hasse diagram $H_{\Cc}$ is a
{\em rooted $X-$forest}\footnote{A rooted $X-$forest $T=(V,E)$ is a collection of rooted trees  together with a function $\phi=\phi_T: X \rightarrow V$ providing a partial labelling of  vertices of $T$ such that every unlabelled  vertex has degree at least $3$.} relative to the labelling map
$\phi_{\Cc}: X \rightarrow \Cc:x\mapsto \Cc(x)$
where the  roots (respectively leaves) are formed by the maximal  (respectively minimal) clusters in $\Cc$ -- see, for example, \cite{dre4,sem} for more details. In particular,
given any  hierarchy $\Cc\in \Cc^{(2)}(X)$ with $X=\bigcup \Cc$, the following holds:
\begin{enumerate}
\item[(H1)]
the associated rooted $X-$forest $H_{\Cc}$ is a rooted $X-$tree (i.e., it is connected as a graph) if and only if $X\in \Cc$ holds.
\item[(H2)]
$\phi_{\Cc}$ maps $X$ bijectively onto the set of leaves of  $H_{\Cc}$ if and only if one has $ \{x\} \in \Cc$ for all $x \in X$.
\item[(H3)]
$H_{\Cc}$ is a rooted binary $X$-tree with leaf set
 $\big\{\{x\}: x\in X\big\}$
 if and only if
$\Cc$ is a maximal $X$-hierarchy (see, for example, \cite{dre4,sem} for details).
\end{enumerate}
Furthermore, associating to every $X$-hierarchy with
$X=\bigcup\Cc$ the corresponding rooted $X-$forest sets up a canonical one-to-one correspondence between such hierarchies and isomorphism classes of rooted $X-$forests, i.e., every
rooted $X-$forest is isomorphic to the rooted $X-$forest associated to one and only one such hierarchy. These concepts are illustrated in Fig.~\ref{fig1} where, for simplicity (here and elsewhere), we write a set  $\{x_1,\dots,x_t\}$ as $x_1\dots x_t$ when no confusion can occur.

\section{Examples of patchworks}\label{examples}

In this section, we collect some examples of patchworks that arise naturally in the study of cluster systems, and present some characterizations of hierarchies via patchworks and adjoint cluster systems.

\begin{enumerate}

\item[(E1)]
Let $\Cc$ be a cluster system for which
$|\bigcup \Cc'| \geq |\Cc'|+p$
holds for some fixed  integer $p$
for all non-empty subsets $\Cc'$ of $\Cc$.
Then the collection of all non-empty subsets $\Cc'$ of $\Cc$ with $|\bigcup \Cc'| =  |\Cc'|+p$ forms a patchwork
since
$$
\big|\bigcup \Cc'\big| +\big|\bigcup \Cc''\big|
= \big|\bigcup \Cc'\cap \bigcup \Cc''\big| +\big|\bigcup \Cc'\cup \bigcup \Cc''\big|
\ge
\big|\bigcup (\Cc'\cap \Cc'')\big| +\big|\bigcup (\Cc'\cup \Cc'')\big|
$$
and
$ \big|\Cc'\big|+\big| \Cc''\big| =\big|\Cc'\cap \Cc''\big|+\big|\Cc'\cup \Cc''\big|$ holds for any two subsets $\Cc',\Cc''$ of $\Cc(X)$ (see also Lemma 1.2 in \cite{dreste}).
In phylogenetic combinatorics, this observation is central to supertree construction from triplet and
quartet trees (with $p=2$ and $p=3$, respectively; see \cite{BDM99,dre4,dreste,sem}).
Moreover, the case $p=0$ has an interesting combinatorial implication for any cluster system $\Cc$ for $X$ that has a `system of distinct representatives'  \cite{hall}:
If $\Cc$ is such a cluster system, then the `trivial' direction of Hall's classic result implies that the collection of non-empty subsets $\Cc'$ of $\Cc$ that
satisfy $|\bigcup \Cc'| = |\Cc'|$ forms a
patchwork (see \cite{Happy} for further related results).

\item[(E2)]
Let $G=(X,E)$ be a finite graph with vertex set $X$ and let  $\Cc_G$ denote the set of subsets of $X$ that are connected relative to $G$. Then, $\Cc_G$ is a weak patchwork. Moreover,  if  $G$ is acyclic (i.e. a forest of trees), then $\Cc_G$ is a patchwork. More generally,
the following can easily be established:
\begin{lemma}
\label{blocklem}
For any finite graph $G$ with vertex set $X$,
$\Cc_G$ is a patchwork if and only if $G$ is a {\em block graph}, i.e., a graph in which every $2$-connected component or `block' is a clique.
\end{lemma}

\item[(E3)]  Further examples of  saturated patchworks are provided by:
\begin{proposition}\label{saturated}
The adjoint $\Cc^*$ of  any cluster system $\Cc\in \Cc^{(2)}(X)$ is a saturated patchwork that contains $\Cc_{triv}$.
\end{proposition}
\pf The fact that $\Cc^*$ contains $\Cc_{triv}$ is obvious -- and follows directly from the observations (F1) and (F3).
The fact that  $\Cc^*$ is saturated follows immediately from the following lemma.
\begin{lemma}\label{Join-compatible}
A cluster $C$ in $\Cc(X)$ is compatible with two given clusters $A,B \in \Cc(X)$ that are not  compatible with each other if and only if either one of the following five assertions holds:

\smallskip
{\rm (i)} $C\cap (A\cup B)=\0,$ \,\,{\rm (ii)}  $A\cup B \subseteq C,$ \,\,
{\rm (iii)} $C  \subseteq A\cap B,$ \,\,{\rm (iv)} $C  \subseteq A-B,$\,\,or 
{\rm (v)} $C  \subseteq B -A.$

\smallskip
In particular, if $C$ is compatible with two clusters $A,B \in \Cc(X)$ that are not  compatible with each other, then the five clusters $A\cup B, A\cap B,  A-B,  B -A,$ and $  (A-B) \cup (B -A)$ must also be compatible with $C$.
\end{lemma}
\pf 
If  $C$ is compatible with two clusters $A,B \in \Cc(X)$ that are not  compatible with each other and (i) does not hold, we may assume wlog that $C\cap A\neq\0$ and, therefore, either $C  \subseteq A$ or $A \subseteq C$ holds. However, if 
 $A \subseteq C$ holds, we must also have $C\cap B\neq\0$ and, therefore,  also $B \subseteq C$ and, hence, $A\cup B \subseteq C$ -- i.e. (ii) -- holds as $C \subseteq B$ cannot hold in view of $A \not\subseteq B$.
 
 Otherwise, we have  $C  \subseteq A$ and, therefore, $B \not\subseteq C$. So, either $C  \subseteq B$ or  $C\cap B= \0$ must hold implying that either $C  \subseteq A\cap B$ -- i.e. (iii) -- or 
 $C  \subseteq A-B$  -- i.e. (iii) --  holds.
 
Conversely, it is obvious that each of these five assertions implies that not only $A$ and $B$, but also the five clusters $A\cup B, A\cap B,  A-B,  B -A,$ and $  (A-B) \cup (B -A)$ must be compatible with $C$. 
\epf

\item[(E4)] Finally, given any cluster system $\Cc\in \Cc^{(2)}(X)$, recall that it was noted in \cite{Elus,IH} that the `ST-sets' with respect to $\Cc$, i.e., the clusters $A$ in its adjoint $\Cc^*$
with $C\!\!\parallel \!\!C'$ for all subsets $C,C'$ of $A$ in $\Cc$, form a weak patchwork.
	Remarkably, this can now be improved in several ways: Consider an arbitrary subset $\Rr$ of the set  $\Cc(X)\times \Cc(X)$ and define, given any such subset $\Rr$ and any cluster system $\Cc\subseteq \Cc(X)$, five subsets of $\Cc^*$ denoted $\Cc^*(\Rr, {\rm sub}), \Cc^*(\Rr, {\rm psub}), \Cc^*(\Rr,  {\rm sup}), \Cc^*(\Rr, {\rm psup}), $ and $\Cc^*(\Rr,  {\rm disj})$ as follows:
\begin{eqnarray*}
\Cc^*(\Rr, {\rm sub})&:=&\{A\in \Cc^*: (C,C')\in \Rr \text{ for all } C,C' \in \Cc  \text{ with } C,C' \subseteq A\},\\
\Cc^*(\Rr, {\rm psub})&:=&\{A\in \Cc^*: (C,C')\in \Rr \text{ for all } C,C' \in \Cc  \text{ with } C,C' \subsetneq A\},\\
\Cc^*(\Rr, {\rm sup})&:=&\{A\in \Cc^*: (C,C')\in \Rr \text{ for all } C,C' \in \Cc  \text{ with } C,C' \supseteq A\},\\
\Cc^*(\Rr, {\rm psup})&:=&\{A\in \Cc^*: (C,C')\in \Rr \text{ for all } C,C' \in \Cc  \text{ with } C,C' \supsetneq A\}, \text{ and }\\
\Cc^*(\Rr, {\rm disj})&:=&\{A\in \Cc^*: (C,C')\in \Rr \text{ for all } C,C' \in \Cc  \text{ with } C\cap A=  C'\cap A=\0\}.
\end{eqnarray*}
Note that 
$$
\Cc^*(\Rr, {\rm sub})\subseteq \Cc^*(\Rr, {\rm psub}) \text{ and }
\Cc^*(\Rr, {\rm sup})\subseteq \Cc^*(\Rr, {\rm psup})
$$ 
always holds, and that $\Cc^*(\Rr, {\rm sub})$ and $\Cc^*(\Rr, {\rm psub})$ contain -- essentially by definition -- all clusters in $\Cc^*$ that are proper sub-clusters of clusters in $\Cc^*(\Rr, {\rm psub})$, i.e.,
\begin{equation}\label{down-closed}
A\in \Cc^*, A' \in \Cc^*(\Rr, {\rm psub}), A \subsetneq A' \Ra A \in \Cc^*(\Rr, {\rm sub})\subseteq \Cc^*(\Rr, {\rm psub})
\end{equation}
holds. Similarly, $\Cc^*(\Rr, {\rm sup})$ and $\Cc^*(\Rr, {\rm psup})$ contain -- also by definition -- all clusters in $\Cc^*$ that  properly contain a cluster in $\Cc^*(\Rr, {\rm psup})$,
 i.e.,
\begin{equation}\label{up-closed}
A\in \Cc^*, A' \in \Cc^*(\Rr, {\rm psup}), A \supsetneq A' \Ra A \in \Cc^*(\Rr, {\rm sup})\subseteq \Cc^*(\Rr, {\rm psup})
\end{equation}
also holds. And so does
\begin{equation}\label{up-closed2}
A\in \Cc^*, A' \in \Cc^*(\Rr, {\rm disj}), A \supsetneq A' \Ra A \in \Cc^*(\Rr, {\rm disj}).
\end{equation}

\smallskip
Furthermore,  given any $\Rr$ as above, let $\Rr_{\rm disj}$ denote the union of $\Rr$ and the set of all pairs $(C,C')\in \Cc(X)\times \Cc(X)$ with $C\cap C' = \0$, let $\Rr_{\!\parallel}$ denote the -- generally larger -- union of $\Rr$ and the set of all pairs $(C,C')\in \Cc(X)\times \Cc(X)$ with $C \!\!\parallel \!\!C'$, and note 
\begin{itemize}
\item[($\parallel$:1)] 
that $\Cc^*(\Rr_{\!\parallel}, {\rm sub})=\Cc^*(\Rr_{\!\parallel}, {\rm psub})$ and
$\Cc^*(\Rr_{\!\parallel}, {\rm sup})=\Cc^*(\Rr_{\!\parallel}, {\rm psup})$ always holds as $C=A$ implies $C\!\!\parallel\!\! C'$ for all clusters $C'\in \Cc(X)$ with either $C'\subseteq A$ or  $C'\supseteq A$, 

\item[($\parallel$:2)]  
and that also $(\Rr_{\!\parallel})_{\rm disj}=\Rr_{\!\parallel}$ and, hence, also $\Cc^*(\Rr_{\!\parallel}, {\rm sub})=\Cc^*(\Rr_{\!\parallel}, {\rm psub})=\Cc^*(\Rr'_{\rm disj}, {\rm psub})$ and
$\Cc^*(\Rr_{\!\parallel}, {\rm sup})=\Cc^*(\Rr_{\!\parallel}, {\rm psup})=\Cc^*(\Rr'_{\rm disj}, {\rm psup})$ holds for $\Rr':=\Rr_{\!\parallel}$.
\end{itemize}

\smallskip
Clearly, the ST-sets from \cite{Elus,IH} are exactly the sets 
$\Cc^*(\0_{\parallel}, {\rm sub})$. Thus, the observation from \cite{Elus,IH}  is implied by

\begin{lemma}\label{ST}
Given any subset $\Rr$ of the set $\Cc(X)\times \Cc(X)$, the following holds:

{\rm (i)} 
The cluster systems  $\Cc^*(\Rr_{\!\parallel},$ ${\rm sub})= \Cc^*(\Rr_{\!\parallel}, {\rm psub})$, $\Cc^*(\Rr_{\rm disj},{\rm psub})$,  $\Cc^*(\Rr_{\!\parallel},{\rm sup})= \Cc^*(\Rr_{\!\parallel}, {\rm psup})$, and $\Cc^*(\Rr_{\rm disj},{\rm psup})$ form saturated patchworks.
 
{\rm (ii)}
The two cluster systems  $\Cc^*(\Rr_{\!\parallel},{\rm disj}) $ and $\Cc^*(\Rr_{\rm disj},{\rm disj})$ form patchworks.
\end{lemma}
\pf (i): In view of ($\parallel$:1) and ($\parallel$:2), it suffices to show that 
the two cluster systems $\Cc^*(\Rr_{\rm disj},{\rm psub})$ and $\Cc^*(\Rr_{\rm disj},{\rm psup})$ form saturated patchworks for any subset $\Rr$ of the set $\Cc(X)\times \Cc(X)$ \big(including those  subsets $\Rr'$ of 
$\Cc(X)\times \Cc(X)$ that are of the form 
$\Rr'=\Rr_{\!\parallel}$ for some subset $\Rr$ of $\Cc(X)\times \Cc(X)$\big).

\smallskip
To do so, recall that, by Lemma \ref{Join-compatible}, either $C\cap (A\cup B)=\0$, or $A\cup B \subseteq C$, or $C  \subseteq A\cap B$, or $C  \subseteq A-B$, or $C  \subseteq B -A$ holds for any cluster $C$ in $\Cc$  and any two clusters $A,B$ in $\Cc^*$ that are not compatible with each other.
Thus, if $C$ is also a proper subset of $A\cup B$, it must be contained in exactly one of the three pairwise disjoint subsets  $A\cap B, A-B,$ or  $B -A$. And if $C$ properly contains either $A\cap B, A-B$, or $B -A$, it must contain $A\cup B$.

 In consequence, $A,B\in  \Cc^*(\Rr_{\rm disj}, {\rm psub})$, 
 $A\!\not\,\parallel \!B$, 
 $C,C' \in \Cc$, and $C,C'\subsetneq A\cup B$ implies that 
either $C,C'\subsetneq A$ or  $C,C'\subsetneq B$ or $C \cap C'=\0$ and, hence in any case, $(C,C')\in \Rr_{\rm disj}$ holds.

And $A,B\in  \Cc^*(\Rr_{\rm disj}, {\rm psup})$, $C,C' \in \Cc$, $A\!\not\,\parallel \!B$ , and $C,C'\supsetneq A\cap B$ or $C,C'\supsetneq A- B$ or
$C,C'\supsetneq B-A$ implies that 
$C,C'\supseteq A\cup B$ and, therefore, also  $C,C'\supsetneq A$ and, hence, again $(C,C')\in \Rr_{\rm disj}$ holds.

So, $A,B\in  \Cc^*(\Rr_{\rm disj}, {\rm psub})$  and $A\!\not\,\parallel \!B$  implies $A\cup B \in \Cc^*(\Rr_{\rm disj}, {\rm psub})$ and, therefore, also 
$A\cap B,  A-B,  B -A,$ 
$(A-B) \cup (B -A)\in  \Cc^*(\Rr_{\rm disj},{\rm psub})$ in view of (\ref{down-closed}). 

And $A,B\in  \Cc^*(\Rr_{\rm disj}, {\rm psup})$ and $A\!\not\,\parallel \!B$ implies $A\cap B,  A-B,  B -A,$ 
$(A-B) \cup (B -A)\in  \Cc^*(\Rr_{\rm disj},{\rm psup})$ while $A\cup B \in \Cc^*(\Rr_{\rm disj}, {\rm psup})$ holds anyway in view of (\ref{up-closed}).

In consequence, the cluster systems $\Cc^*(\Rr_{\rm disj}, {\rm psub})$ and  $\Cc^*(\Rr_{\rm disj}, {\rm psup})$ and, hence, in particular  also  the cluster systems $\Cc^*(\Rr_{\!\parallel},{\rm sub})$ $= \Cc^*(\Rr_{\!\parallel}, {\rm psub})$ and 
$\Cc^*(\Rr_{\!\parallel},{\rm sup})= \Cc^*(\Rr_{\!\parallel}, {\rm psup})$
all are saturated patchworks for every subset 
$\Rr$ of $\Cc(X) \times \Cc(X)$.

\medskip

(ii):  Finally,  Lemma \ref{Join-compatible} implies also that $C\cap (A\cup B)=\0$ or $C  \subseteq A-B$ or $C  \subseteq B -A$ holds for any cluster $C$ in $\Cc$  and any two clusters $A,B$ in $\Cc^*$ that are not compatible with each other and for which $C\cap (A\cap B)=\0$ holds. 

Thus, if $C,C'$ are two clusters with  $C\cap (A\cap B)=C'\cap (A\cap B)=\0$, we have either $C \cap C' = \0$ or $C\cap A =C'\cap A =\0$ or 
$C\cap B =C'\cap B =\0$ and, therefore, in any case  $(C,C')\in \Rr_{\rm disj}$. 

So, $A,B\in  \Cc^*(\Rr_{\rm disj}, {\rm disj})$ and $A\!\not\,\parallel\! B$ implies $A\cap B\in  \Cc^*(\Rr_{\rm disj},{\rm disj})$ while $A\cup B \in \Cc^*(\Rr_{\rm disj}, {\rm disj})$ holds anyway in view of (\ref{up-closed2}).

In consequence, the cluster systems $\Cc^*(\Rr_{\rm disj}, {\rm disj})$ and $\Cc^*(\Rr_{\!\parallel},{\rm disj})$ form patchworks for every subset $\Rr$ of $\Cc(X) \times \Cc(X)$.
\epf

\noindent{\em Remark} We leave it to the interested reader to construct an example of a  cluster system $\Cc\subseteq  \Cc(X)$ for which 
$\Cc^*(\0_{\parallel},{\rm disj})$ does not form a saturated patchwork.
 
 \end{enumerate}


\section{Ample patchworks}\label{ample}

While it is obvious that any cluster system $\Cc\in \Cc^{(2)}(X)$ that contains, for every arc $(C_2,C_1)\in E_{\Cc}$, a  maximal $X$-hierarchy $\Hh$ with $C_1,C_2\in \Hh$ must be ample, it was observed in  \cite{BD-01} that -- somehow conversely -- a patchwork that contains $\Cc_{triv}$ is ample if and only if it contains a maximal $X$-hierarchy, and in  \cite{BS}  that a weak patchwork $\Cc\in  \Cc^{(2)}(X)$  that contains $\Cc_{triv}$ is ample if and only if the following apparently stronger assertion holds:  every hierarchy $\Hh\subseteq \Cc$ can be extended within $\Cc$ to a maximal $X$-hierarchy, i.e., to a maximal $X$-hierarchy $\Hh'\supseteq \Hh$ with 
$\Hh' \subseteq \Cc$. In particular, any such patchwork must contain some maximal $X$-hierarchy. Combining these facts with Proposition \ref{saturated} and our observations collected in (F5), we obtain:

\begin{theorem}\label{motive}
A cluster system $\Cc\in \Cc^{(2)}(X)$ is a hierarchy if and only if
the adjoint $\Cc^*$ of  $\Cc$ is ample  and --
hence -- an ample saturated  patchwork that contains  $\Cc_{triv}$.
Conversely, the adjoint $\Cc^*$ of any ample weak patchwork $\Cc\in  \Cc^{(2)}(X)$ that contains $\Cc_{triv}$ is a hierarchy, implying that its double adjoint
$\Cc^{**}$ must also be ample and -- hence, just as above -- an ample saturated  patchwork that contains  $\Cc_{triv}$.
\end{theorem}
\pf Indeed, if $\Cc$ is a hierarchy, its adjoint $\Cc^*$ contains a maximal hierarchy in view of (F5-ii) and it is a patchwork by Proposition \ref{saturated}. It  must, therefore,  be ample in view of the ``if'' direction of the results obtained in \cite{BD-01}. Conversely, if  $\Cc^*$ is ample, it must -- in view of the fact that it is a patchwork that contains $\Cc_{triv}$ and the
``only if'' direction -- contain a maximal hierarchy. So, its adjoint $(\Cc^*)^*=\Cc^{**}$ and, hence, also $\Cc$ itself must be a hierarchy in view of (F5-i).

The remaining claim regarding the adjoints $\Cc^*$ of ample weak patchworks
follows from the fact that every ample weak patchwork $\Cc \in \Cc^{(2)}(X)$ that contains $\Cc_{triv}$ must, by  \cite{BS}, contain a maximal hierarchy. So, its adjoint $\Cc^*$ must be a hierarchy, again by  (F5-i), and its  double adjoint, therefore,  is an ample saturated  patchwork that contains
$\Cc_{triv}$.
\epf

Theorem \ref{motive} supplies the only equivalence that is not perfectly trivial in the following ten equivalent characterizations of hierarchies:

\begin{proposition}\label{hierar}
 Given any cluster system
$\Cc\in \Cc^{(2)}(X)$, the  following ten assertions are equivalent:
$\bf (i)$  $\Cc$ is a hierarchy,
$\bf (ii)$ $\Cc$ is contained in $\Cc^*$,
$\bf (iii)$ $\Cc$ is contained in $\Cc_\circ$,
$\bf (iv)$ every subsystem $\Cc'$ of $\Cc$ is a saturated patchwork,
$\bf (v)$ every subsystem $\Cc'$ of $\Cc$ is a patchwork,
$\bf (vi)$  every subsystem $\Cc'$ of $\Cc$ is a weak patchwork,
$\bf (vii)$  $\Cc$ is a weak patchwork and
coincides with $\Cc_{extr}$,
$\bf (viii)$ $\Cc_{extr}$ is contained in $\Cc^*$,
$\bf (ix)$ $\Cc^*$ contains a maximal $X$-hierarchy,
$\bf (x)$ $\Cc^*$ is ample.
\end{proposition}

\pf
It is obvious that the implications ``{\bf (i)}$\iff ${\bf (ii)}$\iff ${\bf (iii)}'' as well as
 ``{\bf (i)}$\Rightarrow${\bf (iv)}$\Rightarrow${\bf (v)}$\Rightarrow${\bf (vi)}'' hold and that, in turn, {\bf (vi)} implies that $\Cc$ can not contain any pair $A,B$ of incompatible clusters. So, also ``{\bf (vi)}$\Rightarrow${\bf (i)}'' holds.   Further,
{\bf (i)} implies {\bf (vii)} as {\bf (i)} implies {\bf (vi)} and  $\Cc=\Cc_{extr}$, and
{\bf (vii)} implies {\bf (i)} as the existence of a pair $A,B$ of incompatible clusters in a weak patchwork $\Cc$ would imply that their union $A\cup B$ would be contained in  $\Cc$, but not in $\Cc_{extr}$.

It is also obvious that
{\bf (ii)} implies {\bf (viii)} while, conversely, ``{\bf (viii)} $\Rightarrow$ {\bf (i)}''
holds  because if $\Cc$ were not a hierarchy while {\bf(viii)} holds for $\Cc$, we could choose a minimal pair $A,B\in \Cc$ with $A\!\nparallel\!B$,  (i.e.,
a pair  $A,B\in \Cc$ so that  $A',B'\in \Cc$,
$A'\subseteq A$, $B'\subseteq B$, and $A'\!\nparallel\!B'$ implies $A=A'$ and $B=B'$) while, in view of {\bf(viii)}, we must also have $A\not \in \Cc_{extr}$, that is, $A=A_1\cup A_2$ for some $A_1,A_2\in \Cc$ with
$A_1\!\nparallel\!A_2$. By the minimality assumption, this would imply$A_1\!\!\parallel \!\! B$ and $A_2 \!\!\parallel \!\!B$ and, hence, $A\!\!\parallel \!\! B$
in view of Lemma~\ref{Join-compatible}, a contradiction.

Finally, the equivalence  of {\bf(i)} and {\bf(ix)} has been noted already in (F5-iii), and that of {\bf(ix)} and {\bf(x)} in Theorem~\ref{motive}.
\epf

\noindent{\em Remark}
Although the double adjoint $\Cc^{**}$ of any ample cluster system $\Cc\in  \Cc^{(2)}(X)$ that contains $\Cc_{triv}$ must also be an ample cluster system that contains $\Cc_{triv}$, such a cluster system $\Cc$ does not need to coincide with its double adjoint $\Cc^{**}$ as the example $X:=\{1,2,3\}$ and $\Cc:=\{1,2,3,12,23,123\}$ clearly shows. However, as we shall see in the next section, $\Cc=\Cc^{**}$ will hold in case $\Cc$ is also saturated. Indeed, we will show there that an arbitrary cluster system $\Cc\in  \Cc^{(2)}(X)$ coincides with its double adjoint if and only if it is a saturated patchwork that contains
$\Cc_{triv}$.


\section{A duality theory for cluster systems}\label{dual}

We have noted above that, given a cluster system $\Cc\in \Cc^{(2)}(X)$, the adjoint $\Cc^*$ is a saturated patchwork that contains $\Cc_{triv}(X)$. Here, we will show that, conversely, a cluster system $\Cc\in \Cc^{(2)}(X)$ is of the form $\Cc=\Aa^*$ for some cluster system $\Aa\in \Cc^{(2)}(X)$ whenever -- and, therefore,  if and only if -- it is saturated and contains $\Cc_{triv}(X)$. To this end, we note first:
\begin{lemma}\label{alt}
Given a weak patchwork $\Cc\in \Cc^{(2)}(X)$ and any cluster $A$ in $\Cc$ with at least two distinct maximal proper sub-clusters in $\Cc$ -- i.e., with
$|\Max(\Cc\!:\!A)|\ge 2$ -- either one of the following two mutually exclusive assertions holds:
\begin{itemize}
\item[\rm{[M1-$A$]}]
$\Cc(\subseteq\!A)\!\!\parallel \!\!\Max(\Cc\!:\!A)$ holds -- in particular, any two distinct clusters in $\Max(\Cc\!:\!A)$ are disjoint.
\item[\rm{[M2-$A$]}] $C \cap C'\neq\0$ and $C \cup C' =A$ holds for any two distinct clusters  $C,C'$ in $\Max(\Cc\!:\!A).$
\end{itemize}
Furthermore, if $\Cc$ is saturated, then {\rm{[M1-$A$]}} holds for some cluster $A$ in $\Cc$ with $|\Max(\Cc\!:\!A)|\ge 2$ if and only if one has
$\Max(\Cc\!:\!A)\subseteq \Cc^*$ in which case the cluster system
$$
\langle\Max(\Cc\!:\!A)\rangle:=\{\bigcup \Mm: \0\neq \Mm
\subsetneq \Max(\Cc\!:\!A)\}
$$
must also be contained in $\Cc^*$.
\end{lemma}
\pf Note first that, if $C,C'$ are two distinct clusters
in $\Max(\Cc\!:\!A)$ with $C \cap C'\neq\0$, we must have
 $C \cup C' =A$ as well as  $C \cap C''\neq\0$ for every other cluster $C''$ in $\Max(\Cc\!:\!A):=\Max(\Cc\!:\!A)$.
Indeed, $C\neq C'$ and $C \cap C'\neq\0$ implies $C \cup C'\in \Cc$ and  $C,C'\subsetneq C \cup C' \subseteq A$ and  hence $C \cup C' =A$, as claimed.
 And if $C''$ is yet another cluster
 in $\Max(\Cc\!:\!A)$, we must have
$C''\cap C' \subsetneq C'' =C''\cap A = C''\cap (C \cup C') =(C''\cap C) \cup (C''\cap C')$, and hence $C''\cap C \not =\emptyset$, also as claimed.

 Furthermore, if  \rm{[M1-$A$]} holds, $C$ is any cluster in $\Max(\Cc\!:\!A)$, and $B$ is any cluster in $\Cc(\subseteq\!A)$, then we must have either $B\subseteq C$ or $B\cap C=\0$, as $B\not\subseteq C$ and $B\cap C\neq\0$ would imply that $B'\neq C$ and $B'\cap C\neq\0$ would hold for any maximal cluster $B'\in \Max(\Cc\!:\!A)$ with $B\subseteq B'$. So, \rm{[M1-$A$]} cannot hold in  this case.
This clearly implies the first claim.

To establish the second claim,  assume that
$\Cc\in \Cc^{(2)}(X)$ is a  saturated patchwork, that
$A$\ is a cluster in $\Cc$ with $|\Max(\Cc\!:\!A)|\ge 2$ for which
\rm{[M1-$A$]} holds, and that  $\Mm$ is a non-empty proper subset of
$\Max(\Cc\!:\!A)$. Then, we must also have $M :=\bigcup \Mm \in \Cc^*$ because, if $B$ is any cluster in $\Cc$, we must have $B\!\!\parallel \!\!M$ in case
$B\!\!\parallel \!\!A $, as $A\subseteq B$ implies $M\subseteq B$, $A\cap B=\0$ implies $M\cap B=\0$, and $B\subsetneq A$ implies $B\subseteq M$ or $M\cap B=\0$, depending on whether the unique cluster $C\in \Max(\Cc\!:\!A)$ with $B\subseteq C$ is contained in $\Mm$ or not.  And if  $B\nparallel A$ holds, $B \cap A $ and $A - B$ must be disjoint proper subsets of $A$ in $\Cc$ for which the union is $A$. So, if \rm{[M1-$A$]} holds, we must have
$\Max(\Cc\!:\!A)=\{B \cap A, A - B\}$ and, therefore,  either $\Mm=\{A\cap B\}$ and $M=A\cap B$, or $\Mm=\{ A - B\}$ and $M= A - B$, which implies in both cases that $B \!\!\parallel \!\!M$ holds.   \epf

We are now ready to establish the following theorem:
\begin{theorem}\label{main}
A cluster system $\Cc\in \Cc^{(2)}(X)$ coincides with $\Cc^{**}$  or -- equivalently,\,cf.\,{\rm (F1)}-- is of the form $\Cc=\Aa^*$ for some cluster system $\Aa\in \Cc^{(2)}(X)$
if and only if it is a saturated patchwork that contains
$\Cc_{triv}$.

In particular, $\Hh^{**}=\Hh\cup \Cc_{triv}(X)$ holds for every hierarchy $\Hh\subseteq \Cc(X)$ and, more generally, $\Cc^{**}=\Cc\cup \Cc_{triv}(X)$ holds for every  saturated patchwork $\Cc\subseteq \Cc(X)$, while
$\Hh^{**}=\Hh$ holds for a hierarchy $\Hh\subseteq \Cc(X)$ if and only if $\Hh$ contains $\Cc_{triv}(X)$.
\end{theorem}
\pf  It suffices to show that if $\Cc$ is saturated and contains $\Cc_{triv}(X)$, there exist no clusters in $\Cc^{**}-\Cc$. Otherwise, assume that $B$ is a cluster in $\Cc^{**}-\Cc$ of minimal cardinality. In view of $X\in \CC$, there exists a unique smallest cluster $A$ in $\Cc$ with $B\subseteq A$, {\em viz.} $A:=\bigcap \Cc(\supseteq B)$. If there were two clusters $C_1,C_2\in  \Max(\Cc\!:\!A)$ with $C_0:=C_1\cap C_2\neq \0$ and hence  $C_1\cup C_2=A$ and $C_3:=A- C_0\in \Cc$, we must have $B\not\subseteq C_1, C_2,C_3$ and, therefore,   $|B\cap C_1|, |B\cap C_2|<|B|$ as well as $B\cap C_0\neq \0$
and, therefore,    $B\cap C_1, B\cap C_2\neq \0$.
Thus, $B\in \Cc^{**}$ and
$C_1,C_2\in \Cc \subseteq  \Cc^{**}$ implies
$B\cap C_1, B\cap C_2 \in \Cc^{**}$ in view of Proposition \ref{saturated}.
So, our choice of $B$ implies that $B\cap C_1, B\cap C_2 \in \Cc$ must also hold and, therefore, 
$B=B\cap A=B\cap (C_1\cup C_2)=(B\cap C_1)\cup ( B\cap C_2)\in\Cc$ in view of $(B\cap C_1)\cap ( B\cap C_2)=B\cap C_0\neq \0$, which is a contradiction.

So, by Lemma~\ref{alt}, \rm{[M1-$A$]} must hold and, therefore, also $\bigcup \Mm\in \Cc^*$ for any proper non-empty subset $\Mm$ of $\Max(\Cc\!:\!A)\subseteq \Cc$ while, in view of our assumption
 $\Cc_{triv}\subseteq \Cc$, the subsets in $\Max(\Cc\!:\!A)$ must actually form a partition of $A$.
Thus, any cluster $B\in \Cc^{**}-\Cc$ with $B\subsetneq A=\bigcap \Cc(\supseteq B)$ must be the  union of all those subsets $C$ in $\Max(\Cc\!:\!A)$ with which it has a non-empty intersection (as $C\in \Max(\Cc\!:\!A)$ and, therefore,  $C\in \Cc^*$, $B\in \Cc^{**}$ and $B\cap C \neq \0$ imply $C \subseteq B$ because  $B \subseteq C$ cannot hold by our choice of $A$). That is, we must have
$B\in \langle\Max(\Cc\!:\!A)\rangle\subseteq \Cc^*$ and, hence,
$$
B \in  \langle\Max(\Cc\!:\!A)\rangle \cap \Cc^{**}
\subseteq  \langle\Max(\Cc\!:\!A)\rangle \cap  \langle\Max(\Cc\!:\!A)\rangle^*  =\Max(\Cc\!:\!A)\subseteq \Cc
$$
in view of (F1) and (F3), in contradiction to our assumption $C\not\in \Cc$.
The last remarks follow directly from (F0) and (F3).\epf

Combining Theorem  \ref{main} with our observations in Theorem~\ref{motive}, we can now give a detailed description of the structure of all ample and saturated patchworks that contain $\Cc_{triv}$:


\begin{proposition}\label{hierar*}
A cluster system  $\Cc\in \Cc^{(2)}(X)$ is an ample and saturated patchwork that contains $\Cc_{triv}$ if and only if there exists a $($necessarily unique$)$ hierarchy $\Hh\in  \Cc^{(2)}(X)$ containing $\Cc_{triv}\,(${\em viz.}, the hierarchy $\Hh:=\Cc^*)$ such that $\Cc$ is the disjoint union of the set
$\{X\}$ and the sets $\langle\Max(\Hh:A)\rangle$ where
$A$ ranges over all clusters in $\Hh$ of cardinality at least $2$. In particular:
$$
|\Cc|=1+ \sum_{A\in \Cc^*\!,\, |A|>1}(2^{|\Max(\Hh:A)| }-2)
$$
must hold in this case.
\end{proposition}
\pf We have already noted in Theorem~\ref{motive} that the adjoint $\Cc^*$ of any ample weak patchwork $\Cc\in \Cc^{(2)}(X)$
that contains $\Cc_{triv}$
must be a hierarchy that contains $\Cc_{triv}$. So, writing $\Hh$ for the hierarchy
$\Cc^*$, we must have $\Cc=\Hh^*$ by Theorem \ref{main} in case $\Cc\in \Cc^{(2)}(X)$ is an ample and saturated patchwork that contains $\Cc_{triv}$.
It is also obvious that Lemma \ref{alt} implies that $\langle\Max(\Hh:A)\rangle \subseteq \Hh^* =\Cc$ holds for every cluster $A\in \Hh$ with $|A|>1$ and that the various subsets of $\Cc$ of the form $\langle\Max(\Hh:A)\rangle$ for some such $A\in \Hh$ must be disjoint for  distinct clusters $A$ as, given some non-empty and proper subset $\Mm$ of $\Max(\Hh:A)$, $A$ must be the unique smallest cluster in $\Hh$ with $\bigcup \Mm \subsetneq A$. Further,  $|\langle\Max(\Hh:A)\rangle|= 2^{|\Max(\Hh:A)|} -2$ clearly holds for every $A\in \Hh$ with $|A|>1$.

Thus, it suffices to note that, given any cluster $C\in \Hh^*$with $C\neq X$, one has $C\in \langle\Max(\Hh:A)\rangle$ for that unique smallest cluster $A\in \Hh$ with
$C \subsetneq A$, {\em viz.} the intersection of all cluster $A'\in \Hh$ with
$C \subsetneq A'$. However:
\begin{itemize}
\item[--] $C$ must be compatible with all clusters $C'\in \Max(\Hh:A)\subseteq \Cc$;
\item[--] by construction, $C$ is properly contained in  $A$ but not in any cluster  $C'\in \Max(\Hh:A)$; and
\item[--]  $\Max(\Hh:A)$ must be a partition of $A$ in view of our assumption $\Cc_{triv}(X)\subseteq \Hh$.
\end{itemize}
 So, we have $C'\subseteq C$ for every cluster $C'\in \Max(\Hh:A)$ with $C\cap C' \neq\0$ and, therefore,  -- in view of $C=C\cap A=C\cap \bigcup \Max(\Hh:A) = \bigcup_{C'\in \Max(\Hh:A)} C \cap C' $ -- also
 $C=\ \bigcup \Mm(C)$ for $\Mm(C):=\{C'\in \Max(\Hh:A): C\cap C' \neq\0\}$ which, in turn, implies $\0\neq \Mm(C) \subsetneq \Max(\Hh:A)$ and, therefore, also $C\in \langle\Max(\Hh:A)\rangle$, as claimed. \epf

\section{Generators for patchworks}\label{generators}

Obviously, the intersection $\Cc_1 \cap \Cc_2$ of any two
patchworks $\Cc_1$ and $\Cc_2$ is a patchwork, too.
And the same holds for weak or saturated patchworks.
In addition, the cluster system $\Cc(X)$ is a saturated patchwork and hence is also a (weak) patchwork. Therefore, we can associate, to any cluster  system $\Cc\in  \Cc^{(2)}(X)$, the following patchworks ``generated by $\Cc$'': (i) The unique minimal weak patchwork $\wpg{\Cc}$, (ii)
 the -- generally larger -- unique minimal
patchwork $\pg{\Cc}$,
 and (iii) the -- generally still larger --
unique minimal saturated patchwork $\spg{\Cc}$
that, respectively, contain $\Cc$.
These patchworks are also
called the {\em weak patchwork closure}, the {\em patchwork closure}, and the {\em saturated patchwork closure} of $\Cc$.
Clearly, the corresponding three operators
$P_{wk}, P_{pt}, P_{st}\!: \Cc^{(2)}(X)\ra \Cc^{(2)}(X)$
that map any cluster system $\Cc\in  \Cc^{(2)}(X)$ onto
its weak patchwork closure $\wpg{\Cc},$ its patchwork closure $\pg{\Cc}$ and its saturated patchwork closure $\spg{\Cc}$, respectively,
are `closure operators' on $\Cc^{(2)}(X)$.

Further, we have
 $\Cc\subseteq \wpg{\Cc}\subseteq\pg{\Cc}\subseteq \spg{\Cc}\subseteq\Cc^{**}$
for every $\Cc\in \Cc^{(2)}(X)$. So, $\Aa,\Bb\in \Cc^{(2)}(X)$ and $\Aa \!\!\parallel \!\! \Bb$ imply
not only  $\Aa^{**}\!\!\parallel \!\!\Bb^{**}$ as observed in (F2), but also
$\wpg{\Aa}\!\!\parallel \!\!\wpg{\Bb}$, $\pg{\Aa}\!\!\parallel \!\!\pg{\Bb}$ and $\spg{\Aa}\!\!\parallel \!\!\spg{\Bb}$.

Note further that $\wpg{\Cc\cup \Cc_{triv}}=\wpg{\Cc}\cup \Cc_{triv}$,
$\pg{\Cc\cup \Cc_{triv}}=\pg{\Cc}\cup\Cc_{triv}$, and
$\spg{\Cc\cup \Cc_{triv}}=\spg{\Cc}\cup\Cc_{triv}$ must hold for every cluster system $\Cc\in  \Cc^{(2)}(X)$ in view of (F3) as well as
$\Cc^{**}=\spg{\Cc}\cup \Cc_{triv}$: Indeed, $\spg{\Cc} \subseteq\Cc^{**}$ and, therefore, also 
$\spg{\Cc}\cup \Cc_{triv} \subseteq\Cc^{**}$ must hold in view of
Proposition \ref{saturated} while
$\Cc^{**}\subseteq (\spg{\Cc}\cup \Cc_{triv})^{**}$ must -- in view of $\Cc \subseteq \spg{\Cc}\cup \Cc_{triv}$ -- hold by (F1), and
$(\spg{\Cc}\cup \Cc_{triv})^{**}= \spg{\Cc}\cup \Cc_{triv}$ must hold by Theorem \ref{main}, as $\spg{\Cc}\cup \Cc_{triv}=\spg{\Cc\cup \Cc_{triv}}$ is a saturated patchwork that contains $\Cc_{triv}$.

It is also worth noting that our results imply that the saturated closure
$\spg{\Cc}$ of an ample weak patchwork $\Cc\in \Cc^{(2)}(X)$ that contains $\Cc_{triv}$ is also ample, as $\spg{\Cc}=\Cc^{**}$ must hold and
$\Cc^{**}$ must be ample by Theorem~\ref{motive} as this theorem implies that $\Cc^*$ that must be a hierarchy which, in turn, implies that
$\Cc^{**}=(\Cc^*)^*$ must be ample.

Next, given any cluster system $\Cc\in \Cc^{(2)}(X)$ and any $k\in \N_0$,
we define:
\begin{itemize}
\item[(i) ]
its {\em weak $k$-extension}, denoted
$\ext^{(wk)}_k(\Cc)$,
by setting $\ext^{(wk)}_0(\Cc):=\Cc$ and letting
$\ext^{(wk)}_{k+1}(\Cc)$ denote the union of $\ext^{(wk)}_k(\Cc)$
and the collection of all clusters $A\in \Cc(X)$ that are the union of any two
two incompatible clusters in $\ext^{(wk)}_k(\Cc)$;
\item[(ii) ]
its {\em $k$-extension}, denoted  $\ext^{(pt)}_k(\Cc)$, by setting $\ext^{(pt)}_0(\Cc):=\Cc$, and
letting $\ext^{(pt)}_{k+1}(\Cc)$ denote the collection of all clusters $C\in \Cc(X)$ that are the union or intersection of any two incompatible clusters in $\ext^{(wk)}_k(\Cc)$;
and
\item[(iii) ]
its {\em saturated $k$-extension}, denoted
$\ext^{(st)}_k(\Cc)$, by setting $\ext^{(st)}_0(\Cc):=\Cc$ and letting
$\ext^{(st)}_{k+1}(\Cc)$ denote the union of $\ext^{(st)}_k(\Cc)$ and the
collection of all clusters $C\in \Cc(X)$ for which two incompatible clusters $A,B\in \ext^{(st)}_k(\Cc)$ with $C\in \{A\cup B ,A\cap B,  A-B, B-A, ( A-B) \cup (B-A)\}$ exist.

\end{itemize}

It is also obvious that, for ``$yy$'':=``$wk$'', ``$pt$'' or ``$st$'',
we have:
$$
\ext^{(yy)}_k(\Cc) \subseteq\ext^{(yy)}_{k+1}(\Cc)=\ext^{(yy)}_{1}\big( \ext^{(yy)}_{k}(\Cc) \big).
$$
As one should expect, $k$-extensions can be used to  construct patchwork closures explicitly: \\
\begin{lemma}
\label{lem:ext:gss}
Given any cluster system  $\Cc\in \Cc^{(2)}(X)$, one has:
$$\bigcup_{k} \ext^{(wk)}_{k}(\Cc)=\wpg{\Cc},\,\,\,\,\,\,
\bigcup_k \ext^{(pt)}_{k}(\Cc)=\pg{\Cc}, \,\text{ and }\,\,\,
\bigcup_k \ext^{(st)}_{k}(\Cc)=\spg{\Cc}.
$$
\end{lemma}
\pf
With `$yy$'':=``$wk$'', ``$pt$'' or ``$st$'' as above, we clearly have
$\Cc\subseteq \ext^{(yy)}_1(\Cc) \subseteq \y{\Cc}$ and, therefore,  also  
$\y{\Cc}\subseteq \y{\ext^{(yy)}_1(\Cc)} \subseteq \y{\y{\Cc}}=\y{\Cc}$, i.e., $\y{\Cc}=\y{\ext^{(yy)}_1(\Cc)}$. In consequence, we also have
$\y{\ext^{(yy)}_{k+1}(\Cc)} =\y{\ext^{(yy)}_{1}\big( \ext^{(yy)}_{k}(\Cc) \big)}= \y{\ext^{(yy)}_{k}(\Cc) }$ and hence $\ext^{(yy)}_{k}(\Cc) \subseteq \y{\ext^{(yy)}_{k}(\Cc) }=\y{\Cc}$ for all $k\in \N_0$.

It remains to note that also $\y{\Cc}\subseteq \bigcup_k  \ext^{(yy)}_{k}(\Cc)$ holds, which will follow from noting that $\bigcup_k \ext^{(yy)}_{k}(\Cc)$ is a weak patchwork for ``$yy$'' = ``$wk$'', a patchwork for  ``$yy$'' = ``$pt$'', and a saturated patchwork for  ``$yy$'' = ``$sp$''. Indeed, given $A,B\in \bigcup_k \ext^{(yy)}_{k}(\Cc)$ with $A\!\nparallel\!B $, there exists a smallest natural number $k_0\geq 0$ with $A,B\in \ext^{(yy)}_{k_0}(\Cc)$. So, by construction, we have $A\cup B\in \ext^{(wk)}_{k_0+1}(\Cc) \subseteq \bigcup_k\ext^{(wk)}_{k}(\Cc)$ in case  ``$yy$'' = ``$wk$'', we have $A\cup B ,A\cap B\in\ext^{(pt)}_{k_0+1}(\Cc) \subseteq \bigcup_k \ext^{(pt)}_{k}(\Cc)$ in case ``$yy$'' = ``$pt$'', and we have $, A\cup B, A\cap B, A-B,B-A, A\cup B - A\cap B \in \ext^{(st)}_{k_0+1}(\Cc) \subseteq \bigcup_k \ext^{(st)}_{k}(\Cc)$ in case ``$yy$'' = ``$st$'', as claimed.
\epf

For example, for $X:=\{1,2,3\}$ and $\Cc:=\{12,23\}$, we have $\wpg{\Cc}=\ext^{(wk)}_{1}(\Cc)=\{12,23,123\}$, $\pg{\Cc}=\ext^{(pt)}_{1}(\Cc)=\{2,12,23,123\}$ and $\spg{\Cc}=\ext^{(st)}_{1}(\Cc)=\{1,2,3,12,13,23,123\}.$

\medskip

Remarkably, the clusters in the weak patchwork generated by a cluster system can also be characterized as follows: Given any cluster system $\Cc\in  \Cc^{(2)}(X)$ and a subset $A$ of $X$, the {\em incidence graph} $\Gamma(A;\Cc):=\big(A,E(A;\Cc)\big)$ of $A$ relative to $\Cc$ to be the simple graph with vertex set $A$ and edge set
$
E(A;\Cc):=\big\{\{x,y\} \subseteq {A \choose 2}:\text{ there exists some }
B\in \Cc \text{ with }\{x,y\} \subseteq B \subseteq A  \big\}.
$ Then, we have

\begin{proposition}
\label{thm:ext:gss}
Given any cluster system $\Cc\in \Cc^{(2)}(X)$ and any cluster $A\in \Cc(X)$ with $|A|\ge 2$, the following four assertions are all equivalent:
\begin{itemize}

\item [$\bf (i)$] $A\in \wpg{\Cc}$;

\item [$\bf (ii)$] $\Gamma(A;\Cc)$ is connected;

\item [$\bf (iii)$] there exists some $k< |A|$ and clusters
$A_1, A_2, \dots, A_k\in \Cc$ with

$A=A_1 \cup \cdots \cup A_k$ and $(A_1 \cup \cdots \cup A_{j-1}) \!\nparallel\!A_j$ for all
$j=2, \ldots, k$;
\item [$\bf (iv)$]
$A\in \ext^{(wk)}_{n-1}(\Cc)$.
\end{itemize}

\end{proposition}

\pf
$\bf (i)\Ra \bf (ii)$: This follows from the fact that the cluster system $\{C\in \CC(X):\Gamma(A;\Cc) \text{ is connected}\}$ contains $\Cc$ and is a weak patchwork as $A,A'\in \Cc(X)$ implies that -- in view of $E(A;\Cc)\cup E(A';\Cc) \subseteq E(A\cup A';\Cc)$ -- the graph $\Gamma(A\cup A';\Cc)=\big(A\cup A',E(A\cup A';\Cc)\big)$ is connected provided the two graphs $\Gamma(A;\Cc)$ and $\Gamma( A';\Cc)$ are connected and $A\cap A'\neq \0$ holds.

$\bf (ii)\Ra \bf (iii)$: If $\Gamma(A;\Cc)$ is connected,
we may choose an arbitrary edge
$\{x_1,y_1\}\in E(A;\Cc)$ and find at least some cluster $A_1\subseteq A$ in $\Cc$ with $\{x_1,y_1\}\subseteq A_1$.
And if $A'$ is any subset of $A$ for which clusters $A'_1, A'_2,\dots A'_{k'}\in \Cc$ with $A'=A'_1 \cup \cdots \cup A'_{k'}$ and
$ (A'_1 \cup \cdots \cup A'_{j-1}) \!\nparallel\!A'_j$ for all
$j=2, \ldots, k'$ exist, our assumption that
$\Gamma(A;\Cc)$ is connected implies in case $A'\subsetneq A$ the existence of some edge
$\{x,y\}\in E(A;\Cc)$ with $x\in A'$ and $y \not\in A'$ and, therefore,  some additional cluster $A'_{k'+1}\subseteq A$ in $\Cc$ with $\{x,y\}\subseteq A'_{k'+1}$ and, hence, $A'\!\nparallel\!A'_{k'+1}$. Thus, $A=A_1 \cup \cdots \cup A_{k}$ must hold for any maximal sequence of clusters
$A_1, \dots, A_k \subseteq A$ in $\Cc$ with
$ (A_1 \cup \cdots \cup A_{j-1}) \!\nparallel\!A_j$. Furthermore, we must have $k<|A|$ in this case, as
$ (A_1 \cup \cdots \cup A_{j-1}) \!\nparallel\!A_j$ implies that
$ |A_1 \cup \cdots \cup A_{j-1}| <|A_1 \cup \cdots \cup A_{j-1}\cup A_j|$ for all $j=2, \ldots, k$ and, therefore,
$ |A|=|A_1 \cup \cdots \cup A_k|\ge  1+|A_1 \cup \cdots \cup A_{k-1}| \ge \dots \ge k-1 +|A_1|>k$.

Finally, the implications `$\bf (iii)\Ra \bf (iv) \Ra \bf (i)$' are obvious.
\epf

It is also worth noting that, among all cluster systems that generate a given weak patchwork $\Cc$, there exists 
a unique {\bf minimal} one that we shall also call the {\em base} for $\Cc$, {\em viz.}, the cluster system $\Cc_{extr}$:

\begin{proposition}
\label{prop:base}
Let $\Cc\in  \Cc^{(2)}(X)$ be a weak patchwork. 
Then, one has $\wpg{\Cc'}=\Cc$
for some cluster system $\Cc'\subseteq \Cc$ if and only if  $\Cc'$ contains $\Cc_{extr}$.
\end{proposition}

\pf
This is a direct consequence of the fact that the operator
$P_{wk}: \Cc^{(2)}(X)\ra  \Cc^{(2)}(X)$
satisfies the so-called `anti-exchange axiom' (see \cite{ej}), i.e.,
if $\Cc\in  \Cc^{(2)}(X)$ is a cluster system, $C_1,C_2\in \Cc(X)$ are two distinct subsets of  $X$, and neither $C_1$ nor $C_2$ belongs to $\wpg{\Cc}$, but $C_1$ belongs to  $\wpg{\Cc \cup \big\{\{C_2\}\big\}}$ then $C_2$ does {\bf not} belong to $\wpg{\Cc \cup \big\{\{C_1\}\big\}}$ (as  $C_1\in \wpg{\Cc \cup \big\{\{C_2\}\big\}}- \wpg{\Cc}$ implies $|C_2|<|C_1|$).
\epf

\noindent
In this context, the following is also worth noting:
\begin{itemize}
\item[(1)]
	The analogue of Proposition~\ref{prop:base} does not hold for (proper) patchworks. For example, we have $\pg{\{ 12,14,134\}}=\pg{\{ 12,124,134\}}=\Cc$ for the patchwork $\Cc:=\{1,12,14,124,134,1234\}$, but  $\pg{\{ 12,134\}}=\{1,12,134,1234\}\neq\Cc$ holds for the intersection $ \{ 12,134\}$ of $\{ 12,14,134\}$ and $\{ 12,124,134\}$. The example shows also that the closure operator $P_{pt}$ does not satisfy the anti-exchange axiom in view of $14,124\not\in \pg{\{12,  134\}}=\{1,12,134,1234\}$, but $124\in\pg{\{12,134, 14\}}$ and $14\in \pg{\{12,134, 124\}}$.

Remarkably, it is even possible for a patchwork $\Cc$ to have two disjoint non-empty subsets $\Bb_1, \Bb_2$ with
$\pg{\Bb_1}= \pg{\Bb_2}= \Cc$, but $\pg{\Bb}\neq \Cc$ for every proper subset $\Bb$ of either $\Bb_1$ or $\Bb_2$.
Indeed, if $\Cc$ denotes the collection of all non-empty subsets of $\{1,2,3,4\}$ that contain the element $1$, then $\Cc$ forms a patchwork for which $\pg{\Bb_1}= \pg{\Bb_2} = \Cc$ holds for $\Bb_1: = \{12, 13, 14\}$ and $\Bb_2: = \{123, 124, 134\}$, but no proper subset of either $\Bb_1$ or $\Bb_2$ generates $\Cc$.

\item[(2)]
Obviously, the base $\Cc:=\Cc(X)_{extr}$ of $\Cc(X)$ consists of all clusters $C\subseteq X$ of size 1 and 2.
This shows that the size of $\wpg{\Cc}$ can be exponential compared to that of $\Cc$.

\item[(3)]
It is also possible for the base of a weak patchwork $\Cc\in  \Cc^{(2)}(X)$  to be exponential in the cardinality of $X$. For example, if $n$ is even, let $\Cc$ be the set
of all subsets of $X$ of size at least $\frac{1}{2}n$. Then,
the base $\Cc_{extr}$ of $\Cc$ coincides with the set
of all subsets of $X$ of size exactly $ \frac{1}{2}n$,
and its cardinality is exponential in $n$.

\end{itemize}

\section{An illustrative analysis of a biological data set} \label{real-world}

\begin{figure}[htb]
\centering

\xymatrix@!=0.2pc{
&&&&&&&&F_{02} \ar@{-} [d]\ar@{-} [rr]&&G_{02} \ar@{-} [d]
\\
&&&&&\circ\ar@{-} [drr]\ar@{-}[ld]\ar@{-}[rd]&&&\circ\ar@{-} [rd]\ar@{-} [ld]\ar@{-} [rr]&& \circ \ar@{-} [rd]
\\
&&&&A_{00} \ar@{-} [rd]\ar@{-} [drr]&&A_{96}\ar@{-}[ld]
\ar@{-} [drr]&\circ \ar@{-}[ld]\ar@{-}[rd]&&\circ \ar@{-} [ld]\ar@{-} [rr]\ar@{-} [d]&& \circ\ar@{-} [d]
\\
&&&&&\circ\ar@{-} [drr]&\circ\ar@{-}[rd]
&&\circ\ar@{-} [d]\ar@{-} [ld]&\circ \ar@{-} [ld]\ar@{-} [rr]\ar@{-} [d]&&B_{96}\ar@{-} [d]
\\
&&&&&&& \circ\ar@{-}[d]&\circ\ar@{-} [d]\ar@{-} [ld]&\bullet \ar@{-} [ld]\ar@{-} [rr]&&\circ \ar@{-} [rr]  \ar@{-} [ld]&&C_{02},E_{01}
\\
&&&&&&& \circ\ar@{-}[d]&\circ\ar@{-} [rr]\ar@{-} [ld]&&\circ\ar@{-} [ld]
\\
&&&&&&& \circ\ar@{-}[d]\ar@{-} [rr]&& \circ\ar@{-}[d]
\\
&&&&&&&B_{90}\ar@{-} [rr]&&H_{98}}
\caption{The splits graph for the non-trivial splits in  $\Sigma_{HIV}$. Here ``$\bullet$'' stands for the three sequences $C_{94},D_{01}$ and $I_{99}$.}
\label{fig2}
\end{figure}
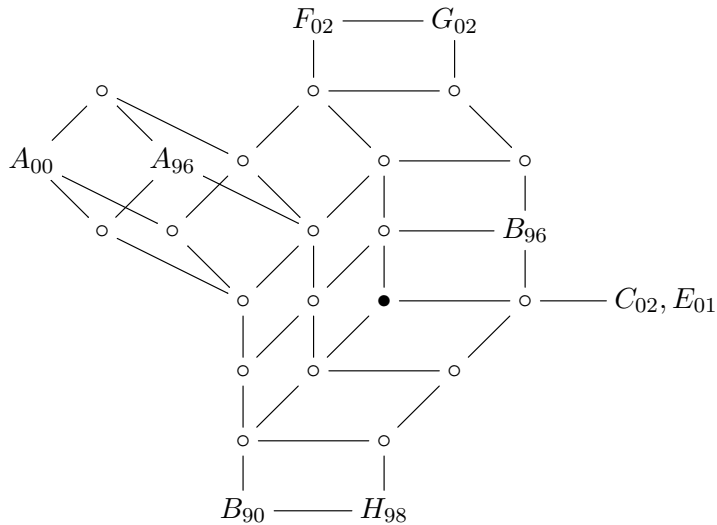

To illustrate our results, we now present
a simple application to a `real-world' data set:
the so-called `Belgian Transmission Chain' of the human immunodeficiency virus 
studied by Lemey {\em et al.} in  \cite{lem}. The original 
data set $X:=\{A_{96}, A_{00}, B_{90}, B_{96}, C_{94}, C_{02}, D_{01}, E_{01}, F_{02}, G_{02},H_{98}, I_{99}\}$ was downloaded from  \cite{hiv}. It contains $12$ env-gp41 HIV sequences from nine patients. The letters indicate the nine distinct patients 
$A,B,\dots, I$ and the indices give the date of isolation. After aligning the sequences, eliminating all sites that contained indels or nucleotides
that could not clearly be identified as either purine (R) or pyrimidine (Y), and then rewriting the sequences simply as $R,Y$-sequences, we obtained an alignment of $12$ sequences with  altogether $902$ sites of which
$875$ were constant while the remaining $27$ non-constant sites induced a split system\footnote{As usual (see, for example,  \cite{dre4,sem}), we call  a bipartition or {\em split} $S=\{A,B\}$ of the set $X$ into two non-empty disjoint subsets $A,B$ of $X$ to be a {\em $k$-split} if $\min(|A|,|B|)=k$ holds, and {\em trivial} if it is a $1$-split. Two splits $S$ and $S'$ of $X$ are called {\em compatible} if  $A\cap A'=\0$ holds for some $A\in S$ and some $A'\in S'$. A collection of splits of $X$ -- or a {\em split system} -- is called {\em compatible} if any two splits in that collection are compatible. And it is called {\em weakly compatible} if  one of the four intersections $A\cap A'\cap A'', A\cap (X-A')\cap (X-A''),  (X-A)\cap A'\cap (X-A''),$ and $(X-A)\cap (X-A')\cap A''$ is empty for any three splits  $S,S',S''$ in that collection and all $A\in S, A'\in S'$ and $A''\in S''$.} 
 $\Sigma_{HIV}$ containing seven distinct trivial splits
 of total multiplicity $14$, four distinct $2$-splits of total multiplicity $7$, two distinct $3$-splits of multiplicity $1$ each, one $4$-split of multiplicity $1$, and two distinct $5$-splits of total multiplicity $3$.
A corresponding splits graph ({\em cf.} Chapter 4.4 in \cite{dre4}) is depicted in Figure~\ref{fig2} in which, for clarity of presentation, all seven $1$-splits separating each one of the seven sequences $A_{96}, A_{00},  B_{96}, C_{02}, D_{01}, F_{02}, $ and $I_{99}$ from the other $11$ sequences are omitted,
and edge lengths are chosen only to avoid ambiguous overlapping and do not indicate `biological weight' or multiplicity.

\bigskip

\bigskip\noindent
We then formed $24$ distinct cluster systems: $12$ by forming, for every sequence $x\in X$, the  cluster system that contains all `split halves' that contain $x$, and 12 that contain exactly all the complements of these split halves. We  then computed their adjoints and their double adjoints. As was to be expected\footnote{As even for a `compatible split system' corresponding to a phylogenetic tree, there would be lots of overlap.}, the adjoints and the double adjoints computed for the first $12$ cluster systems were rather trivial; the adjoints either consisted of $\Cc_{triv}(X)$ or, in three cases, contained one additional cluster of  cardinality $2$,  implying that the double adjoints either coincide with
$\Cc(X)$ and, thus, have cardinality $4049=2^{12}-1$ or -- again in  those three cases -- with a subset of $\Cc(X)$ of cardinality $2049$. The adjoints and the double adjoints computed for the other $12$ cluster systems are more interesting: Ten of the adjoints have cardinality $19$, one adjoint associated with $E_{01}$ has cardinality $21$ and the largest one associated with $C_{94}$ has cardinality $25$ while ten of the double adjoints have cardinality $262$ and those associated with $E_{01}$ and $C_{94}$ have cardinality $261$. None of them is a hierarchy, and all but one of the hierarchies obtained as intersections of the adjoints and the double adjoints contain $16$ clusters, that is, three more than there are clusters in $\Cc_{triv}(X)$. Furthermore, either the cluster $\{C_{02},E_{01}\}$ or $X-\{C_{02},E_{01}\}$ and either $\{C_{94},D_{01},I_{99}\}$ or $X-\{C_{94},D_{01},I_{99}\}$ turn up invariably.

So, while this does not provide too much information (as had to be expected in view of the fact that the split system  $\Sigma_{HIV}$ is not even weakly compatible), it is remarkable that our analysis does not only strongly support the cluster $\{C_{02},E_{01}\}$ that is also supported by all other methods we investigated, but also the cluster  $\{C_{94},D_{01},I_{99}\}$ even though it is generally not supported by other methods and the corresponding split $\{C_{94},D_{01},I_{99}\}|X-\{C_{94},D_{01},I_{99}\}$ is not even one of the  $16$ splits induced by the $27$ non-constant sites of the input alignment.

In a subsequent paper dealing with applications, we will discuss this and related phenomena and their possible biological significance in more detail.

\section{Concluding remarks}  \label{conclusion}

\begin{itemize}
\item[(1)]
The observation that $\Cc_\circ$ is a hierarchy for every  cluster system  $\Cc\in\Cc^{(2)}(X)$ allows us to associate a canonical hierarchy to an arbitrary cluster system that can actually be further enlarged -- but not canonically -- by adding any arbitrary subset in $\Cc$ or in $\Cc^*$, or even by forming the union $\Cc_\circ\cup \Cc'\cup \Cc''$ where $\Cc'$ is  any arbitrarily chosen hierarchy in $\Cc^{**}$ and $\Cc''$ is an arbitrarily chosen hierarchy in $\Cc^*$.
This generalizes a construction described in  \cite{dre} where we discussed the problem of relating, at least on a purely theoretical level, species trees to genealogical history of individual organisms. It was the starting point for the investigations presented here, and it might be of interest to relate the discussion in \cite{dre} and related discussions in  \cite{ald,baum,mat} to the constructions presented here.

\item[(2)]Next, we briefly outline two possible phylogenetic
applications of some of our results (for conciseness, we refer readers who are unfamiliar with some of the phylogenetic terminology to
\cite{dre4,sem}):

For the first application, suppose we have a collection $\Rc$ of rooted
phylogenetic trees having leaf sets that comprise subsets of $X$.
We use the symbol `$ab|c$', for any three distinct elements $a, b, c \in X$, as a 
shorthand for the  `rooted triplet' formed by $a,b$ `versus' $c$, i.e., the  pair $(\{a, b\}, \{c\})$ of subsets of $X$ consisting of the $2$- subset $\{a, b\}$ and the $1$- subset $\{c\}$ of $X$. Using this notation, consider the set $\Rc_3$ of `rooted triplets'  $xy|z$ `displayed' by the trees in  $\Rc$, i.e., all triplets $xy|z$ for which some tree $T_0\in \Rc$ and some edge $e_0$ in $T_0$ exist such that 
\begin{itemize}
\item[(i)] 
$x,y$, and $z$ are leaves of $T_0$ and
\item[(ii)]  $e_0$ separates $x$ and $y$ from $z$ and the root of $T_0$. 
\end{itemize}
Now,
for real data, $\Rc_3$ will typically be `incompatible' (i.e. no tree will
display all the rooted triplets in $\Rc_3$). However, we might hope that
$\Rc_3$ is `sufficiently
comprehensive', i.e., 
that some (unknown) subset $\Rc_3'$ of
$\Rc_3$ `defines' some hopefully true species tree $T$ with leaf set $X$, in the
sense that $T$ is the only tree that displays the rooted triplet trees
in $\Rc_3'$. In this case, it is well known that $T$ must be a `fully
resolved' tree (i.e., the `clades' of $T$, i.e., the sets of leaves that can be separated from its root by some edge, must form a maximal $X$-hierarchy $\Cc_T$).
In general, the collection of all clusters present in the trees in the original
collection $\Rc$ may fail to contain the clusters of $T$. However, the weak
patchwork closure provides a formal way to obtain a superset of $\Cc_T$ from $\Rc_3$ and thus, at least, captures all of its clades.
More precisely we have the following result:
\begin{proposition}
If $\Rc_3$ contains a set of rooted triplets from $X$ that defines a
rooted phylogenetic tree $T$ on $X$, then $\Cc_T$ is a subset of 
$\wpg{\Cc_{\Rc_3}}$ where $\Cc_{\Rc_3} :=
\{\{x,y\}: \exists_{z\in X}xy|z \in \Rc_3\}.$
In particular, the adjoint $\Cc_{\Rc_3}^*$ of $\Cc_{\Rc_3}$ must be a hierarchy that is contained in  $\Cc_T^*= \Cc_T$, and its double adjoint must be ample.
\end{proposition}
The proof of this result in the special case where $\Rc$ itself defines
$T$ uses induction on the height of $T$, together with the well-known
result that, if $\Rc$ defines $T$, then each interior edge of $T$ must be
`distinguished' by at least one rooted triplet from $\Rc$. The general
case where $\Rc$ merely contains some subset
$\Rc_3'$ of rooted triplets
that defines $T$ follows immediately, since $\wpg{\Cc_{\Rc_3'}} \subseteq
\wpg{\Cc_{\Rc_3}}$ holds for every subset $\Rc_3'$ of $\Rc_3$.

As a second possible application, suppose that (i) $T$ is a rooted binary
phylogenetic tree with leaf set $X$
and (ii) $T'$ is some `perturbation' of this
tree obtained by applying some subtree rearrangement operation to $T$
(i.e., re-attaching some subtree of $T$ to a different part of the tree,
as described further in \cite{sem}). Let $\Cc$ be the union of the
clusters of
$T$ and $T'$. Then, if the tree rearrangement corresponds to a `nearest
neighbour interchange' (NNI), one has $\wpg{\Cc}= \Cc$. However, if the
rearrangement corresponds to moving one branch of $T$ further across $T$
than an NNI move allows, the set $\wpg{\Cc}$ will, in general, be
considerably larger than $\Cc$. The former observation can be extended to allow
more than one perturbation. In particular, if the NNI moves occur at
`well separated' nodes of $T$, the union of the clusters of $T$, along
with the clusters of all the resulting perturbed trees will
form a weak patchwork. In this
way, weak patchwork closures may be a possible tool for helping to
distinguishing local errors (due to lack of phylogenetic signal, or
lineage sorting) from more extreme rearrangements events, as might occur
with gene trees in settings where some lateral gene
transfer events have occurred between distantly related taxa.

\item[(3)]
Assume that $\Cc$ is a weak and  ample patchwork that contains $\Cc_{triv}$. In view of Theorem~\ref{motive}, this implies that the hierarchy $\Cc_\circ$ is contained in a maximal $X$-hierarchy that is contained in $\Cc$. 
However, it is easy to see that $\Cc_\circ$ need not be a maximal $X$-hierarchy itself, and that  many distinct maximal $X$-hierarchies may contain $\Cc_\circ$. For example, $\Cc(X)$ is an ample patchwork, yet $\Cc(X)_\circ$ consists of the trivial clusters only. So, it does not seem to be always possible to find a maximal $X$-hierarchy in an ample patchwork in a `canonical' way.

\item[(4)] It should also be of some interest to work out the corresponding theory for split systems rather than cluster systems, as well as for {\em weighted} cluster and split systems. For example, denoting
\begin{itemize}
\item[(i)] 
by $\|C\|:=|\{C\}^*|=2^{|C|} + 2^{|X-C|+1}-2$, for any cluster 
$C\in \Cc(X)$, the number of 
clusters $C'\in \Cc(X)$ that are compatible with $C$, 
\item[(ii)]
by $|\varphi|$, for any map 
$\varphi:\Cc(X)\ra \R$, the $\ell_1$-norm 
$\sum_{C\in \Cc(X)}|\varphi(C)|$ of $\varphi$,
\item[(iii)]
and the set of all {\em weighted cluster systems} for $X$ of  $\ell_1$-norm $1$ by 
$$
\Ww_1(X):=\{\varphi\in \R_{\ge 0}^{\Cc(X)}: |\varphi|=1\},
$$
\end{itemize}
one could study the map
$$
\Ww_1(X)\ra \Ww_1(X):\varphi\mapsto \varphi^*
$$
where $\varphi^*(C)$ is defined, for any map $\varphi\in \Ww_1(X)$ 
and any cluster $C\in \Cc(X)$ by, say,
$$
\varphi^*(C):=\frac{\sum_{C'\in \{C\}\!^* }\varphi(C')}{\sum_{C\in \Cc(X)}
\|C\|\varphi(C)},
$$
a map that maps the $(2^{n}-2)$-dimensional simplex $\Ww_1(X)$ indeed into itself in view of
$$
\sum_{C\in \Cc(X) }\sum_{C'\in \{C\}\!^* }\varphi(C')=\sum_{C'\in \Cc(X)}
\|C'\|\,\varphi(C').
$$

\item[(5)]
Several interesting questions remain:

-- Is there another way to describe or characterize  $\Cc_\circ$? In particular, can it be calculated in polynomial time in $n$ and $|\Cc|$?

-- Does $4n-2\le |\Cc^*|+|\Cc^{**}|\le 2^n+n$ hold for every
non-empty cluster system $\Cc\subseteq \Cc(X)$, i.e., are the cases where $\Cc$ is a maximal $X$-hierarchy and $\Cc=\Cc(X)$ the `extremal' cases when asking for the minimum and the maximum of $ |\Cc^*|+|\Cc^{**}|$ over all non-empty cluster systems $\Cc\subseteq \Cc(X)$?

-- As $\bigcup_k \ext^{(st)}_{k}(\Cc)=\spg{\Cc}$ is ample whenever $\Cc$ is an ample weak patchwork that contains $\Cc_{triv}$, does the same hold also for $ext^{(st)}_{1}(\Cc)$, i.e., is $ext^{(st)}_{1}(\Cc)$ ample whenever $\Cc$ is an ample weak patchwork that contains $\Cc_{triv}$? If yes, some rather detailed case-by-case considerations seem to be required to establish this fact.

-- Also, Proposition~\ref{thm:ext:gss}  tells us that $\wpg{\Cc} =  \ext^{(w)}_{n-1}(\Cc)$ always holds, and it is natural to ask whether an analogous result holds for the other two patchwork-closure operators, too. In other words, can we find  some less than exponentially growing integer function $f=f(n)$ with $\pg{\Cc} =  \ext^{(pt)}_{f(n)}(\Cc)$ or $\spg{\Cc} =  \ext^{(st)}_{f(n)}(\Cc)$ for all cluster systems $\Cc\subseteq \Cc(X)$ and, if so, how slowly can $f$ grow as a function of $n$?

\item[(6)] In view of the investigations presented in \cite{ddggr}, it may also be of interest to replace the binary relation `$\parallel$' by any of the {\em relaxed} relations `$\parallel_k$' ($k\in \N$) defined by:
$$
A \!\parallel_k \!B \iff  |A \cup B| \le \max(|A| ,|B|) + k
$$
for all $A,B\in \Cc(X)$ and to study the associated `duality theory'.
\end{itemize}
\bigskip

{\bf Acknowledgements}

\medskip
AD  thanks the Chinese Academy of Sciences and the Max Planck Society,
MS thanks the Royal Society of NZ (Marsden Fund and James Cook fellowship) and  TW is grateful for
support from the
Singapore MOE grant R-146-000-134-112.  \\


\end{document}